\documentclass[showpacs,preprintnumbers,amsmath,amssymb, 
eqsecnum,
twocolumn, tightenlines,
]{revtex4}

\usepackage{bm}

\usepackage{amsmath}

\usepackage{dcolumn}

   \usepackage[dvips]{graphicx}
\begin{document}

    \bibliographystyle{apsrev}
    
    \title {Reflection on event horizon for collapsing black holes}
    
    \author{M.Yu.Kuchiev}
    \email[Email:]{kuchiev@newt.phys.unsw.edu.au}
    
    \affiliation{School of Physics, University of New South Wales,
      Sydney 2052, Australia}
    
    \date{\today}

    \begin{abstract}
      It was argued recently that there exists an unexpected
      phenomenon, the reflection of incoming particles on the event
      horizon of black holes (Kuchiev(2003)). This means that a
      particle approaching the black hole can bounce back into the
      outside world due to those events that take part strictly on the
      horizon. Previously the effect was discussed in relation to
      eternal black holes. The present work shows that the effect
      exists for collapsing black holes as well.
    \end{abstract}


    \maketitle
    
    \section{introduction}
    \label{intro}
    
    An interest in quantum properties of the scattering problem that
    describes an impact of a particle on a black hole was inspired by
    the discovery of the Penrose process \cite{penrose_69}, and the
    works of Zel'dovich \cite{zeldovich_71_72} and Misner
    \cite{misner_72} devoted to the energy extraction from the Kerr
    black hole. First numerical results for scattering of scalar
    particles on rotating black holes were presented by Press and
    Teukolsky \cite{press_teukolsky_72,press_teukolsky_74}. The
    analytical solution of the scattering problem, which had a strong
    influence on further developments, was found by Starobinsky
    \cite{starobinsky_73} for the scalar field and Starobinsky and
    Churilov \cite{starobinsky_churilov_73} for electromagnetic and
    gravitational waves scattered by the rotating Kerr black hole.
    Independently, Unruh \cite{unruh_76} considered scattering of
    scalar and fermion particles on Schwarzschild black holes. Further
    study of the scattering problem was presented by Sanchez
    \cite{sanchez_1977}.  After these and the numerous works that
    followed, it has been assumed that the scattering problem is
    completely understood, see details and bibliography in
    the books \cite{frolov_novikov_98,thorne_1994,chandrasekhar_93,%
      fullerman_handler_matzner_88}.
    
    The present work discusses particular aspects of the scattering
    problem related to quantum phenomena, which take place strictly on
    the event horizon of black holes. The important relevant effects
    in this area are the Hawking radiation
    \cite{hawking_74,hawking_75}, and the Unruh process
    \cite{unruh_prd_76}.  The Hawking radiation is associates with the
    entropy of black holes, which, according to Bekenstein
    \cite{bekenstein_1972,bekenstein_1973,bekenstein_1974}, equals the
    area of the horizon. For a review of the Hawking radiation and the
    thermodynamics properties of black holes see Refs.
    \cite{jacobson_03} and \cite{wald_2001}.  The Hawking radiation
    can be treated as a tunneling process, see e.g. Parikh and Wilczek
    \cite{parikh_wilczek_00}, Khriplovich \cite{khriplovich_98}, and
    Khriplovich and Korkin \cite{khriplovich_korkin_02} and references
    therein.  Another line of research on quantum phenomena aims at
    quantization of energy levels of black holes, see discussion in
    Bekenstein \cite{bekenstein_2002_a,bekenstein_2002_b} and
    references therein.  Treatment of a number of different quantum
    phenomena on the event horizon can be found in the review by 't
    Hooft \cite{thooft_96} which, among other issues discusses the
    brick wall model of Ref.  \cite{thooft_85}, aimed at explanation
    of the entropy of black holes.
 
    The effect of the Hawking radiation can be considered in terms of
    the influence that the horizon exercises on the vacuum of the
    theory.  It has been realized recently that quantum events on the
    horizon have also a strong impact on the wave function of a
    particle that approaches a black hole
    \cite{kuchiev_1,kuchiev_2,kuchiev_3}. As a result there arises a
    new qualitative feature in the scattering problem. A particle
    approaching the black hole can bounce on its event horizon back
    into the outside world.  This phenomenon, which is referred to
    below as the reflection on the horizon (RH) using the terminology
    of \cite{kuchiev_2}, is due to pure quantum reasons; in the
    classical approximation it is obviously absent.  The RH effect,
    which is strong when the wavelength of the particle exceeds the
    radius of the event horizon, reduces the absorption cross section,
    making it zero in the infrared region, as was demonstrated
    explicitly in Ref.\cite{kuchiev_flambaum_04} for scattering of
    low-energy scalar massless particles on Schwarzschild black holes.
    This behavior of the cross section differs qualitatively from the
    result of Unruh \cite{unruh_76}, which stated that this cross
    section equals the area of the horizon.  Ref.
    \cite{kuchiev_flambaum_04} argued that similar reduction of the
    cross section is expected for scattering of any massless particle
    by any black hole in the low energy limit.  Recent Ref.
    \cite{kf_review} gives a brief summary of progress related to the
    RH.

    Refs.\cite{kuchiev_1,kuchiev_2,kuchiev_3,kuchiev_flambaum_04,kf_review}
    suggested several arguments in favor of the RH.  However, all of
    them, one way or another, were related to eternal black holes,
    inspiring a question as to whether the phenomenon exists for
    collapsing black holes as well, or is merely a peculiar feature
    relevant to eternal black holes only.  The present work shows that
    the RH does take place for collapsars being thus a common effect
    for black holes.

    \section{Symmetry and analytical continuation}
    \label{main}
    Let us formulate briefly and in general terms the main idea of
    this work. Consider a quantum system characterizes by a set of
    coordinates $Q$, which include the time variable; generally
    speaking this set may be infinite. Take some quantum state of this
    system, not necessarily stationary, which is described by the wave
    function $\Psi(Q)$. Assume now that there is some discrete
    symmetry that characterized the system, calling the operator that
    generates this symmetry $\hat S$. It suffices to restrict out
    presentation to the simplest case, when the symmetry group is
    $Z_2$, i.e. the symmetry group is generated by the only operator
    $\hat S$, which satisfies $\hat S^2 =1$.  (Though more general
    symmetries can also be covered by the method discussed, they will
    not play a role in the present work.)  Since $\hat S$ is a
    symmetry, the function $\Psi' (Q)=\hat S[\Psi](Q)$ necessarily
    describes some physically allowed state of the system.
    
\begin{figure}[tbh]
      \centering
      \includegraphics[height=2.5cm,keepaspectratio=true]{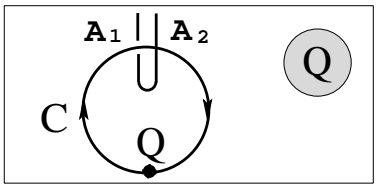}
      \caption{Schematic presentation of the analytical continuation of 
        the wave function $\Psi(Q)$ into the complex-valued
        coordinates $Q$. The contour $C$ goes from the sheet $A_1$ of
        the Riemann surface over the cut to another sheet $A_2$. As a
        result the wave function changes from $\Psi(Q)$ into $\tilde
        \Psi(Q)$.}
      \label{zero}
    \end{figure}

    Let us consider now the complex extensions of the coordinates $Q$,
    using the analytical continuation for constructing the wave
    function $\Psi(Q)$ for the complex-valued coordinates. Take some
    closed contour $C$ that runs in the multidimensional space of the
    complex coordinates $Q$. Fig.  \ref{zero} gives a schematic
    presentation for this contour. The contour starts and finishes at
    some real physical point $Q$.  Generically, the wave function
    $\Psi(Q)$ should have a sufficiently complex Riemann surface that
    possesses some singularities; the cuts attached to them give
    access to different sheets of the Riemann surface.
    Correspondingly, the value of the wave function depends on the
    sheet of the Riemann surface where it is taken. Thus, if the
    contour $C$ crosses some cut that separates two sheets of the
    Riemann surface, called $A_1$ and $A_2$ in Fig.  \ref{zero}, then
    after returning to the initial physical value of $Q$ the wave
    function acquires a new value, call it $\tilde \Psi(Q)$.
    Assume now that one is able to verify that the analytical
    continuation along the contour $C$ and the symmetry transformation
    with the operator $\hat S$ have one and the same influence on the
    system. Then one can say that
    \begin{eqnarray}
      \label{psipsi}
      \hat S[\Psi] (Q)=\tilde \Psi(Q)~. 
    \end{eqnarray}
    This equality is derived from the initial wave function $\Psi(Q)$
    by the symmetry transformation, which is implemented on the
    left-hand side directly, while on the right-hand side it works
    through the analytical continuation. One concludes from this that
    Eq.(\ref{psipsi}) provides a symmetry condition on $\Psi(Q)$.
    Eq.(\ref{psipsi}) is based on the relation between a monodromy
    transformation (analytical continuation) and a discrete symmetry
    of the system, which generally speaking is a well known method. To
    mention some result derived with its help one can remember the
    known Seiberg-Witten solution for the $N=2$ supersymmetric theory
    Ref.\cite{seiberg-witten_94}.
    
    The system discussed in this work consists of collapsing matter
    and an incoming probing particle.  The relevant discrete symmetry
    of the system is the time inversion, which transforms the
    collapsing matter (black hole) into an expanding matter (white
    hole), and at the same time forces the incoming particle to become
    the outgoing one. We will see in Section \ref{wave} that
    Eq.(\ref{psipsi}) puts a restriction on the wave function of the
    particle, deriving from it that the RH really takes place.

    \section{Collapsing black hole}
    \label{tolman}
    This section summarizes briefly several important for us, though
    well known facts related to the collapse of the dust matter, i.
    e.  matter without pressure, $p=0$, see books
    \cite{misner_thorne_wheeler_1973,landau_lifshits_II} for details.
    For simplicity let us take the spherically symmetrical case when
    the collapse can be described by the Tolman metric
    \begin{eqnarray}
      \label{metric}
      ds^2 = -d\tau^2+
      \exp(\lambda)\,dR^2+r^2d\Omega^2~,
    \end{eqnarray}
    which is a function of $\tau$, the radial variable $R$, and angular
    variables $\theta,\varphi$; $d\Omega^2= d\theta^2+\sin^2\theta
    d\varphi^2$.  The functions $\lambda=\lambda(R,\tau)$ and
    $r=r(R,\tau)$ are defined by the Einstein equations, which give
    for them
    \begin{eqnarray}
      \label{lam}
      &&      \exp(\lambda)=\frac{r'^{\,2}}{f(R)+1}~, \\ 
      \label{rdot}
      && \dot{r}^{\,2}=f(R)+F(R)/r~.
    \end{eqnarray}
    Here and below the dotted and primed functions indicate
    derivatives over $\tau$ and $R$.  The two functions $f(R),~F(R)$
    provide the initial conditions for the spherically symmetrical
    dust matter.  One of them, namely $F(R)$, defines the density of
    matter $\epsilon$
    \begin{eqnarray} 
      \label{eps}
      8\pi G\epsilon = \frac{F'}{r'r^2}~.
    \end{eqnarray}
    The total mass of matter inside a sphere with the radius $R$
    equals
    \begin{eqnarray}
      \label{mass}
      m(R)= 4\pi \int_0^R \epsilon \,r^2 \,dr=\frac{F(R)}{2G}~.
    \end{eqnarray}
    The integrand here has no additional factors of $\exp(\lambda)$ or
    $r'$ due to the defect of mass caused by gravity.  Eq.(\ref{mass})
    shows that $F(R)$ can be considered as the gravitational radius
    for the mass of matter accumulated inside the sphere of radius
    $R$, $F(R) = 2\,G\,m(R)$.  Another function $f(R)$, which appears
    in the formulation of the problem, gives an additional parameter
    that is needed to define the distribution of velocities of the
    matter given by Eq.(\ref{rdot}).  The two functions $f(R),F(R)$
    will be presumed to satisfy conditions
    \begin{eqnarray}
      \label{condi1}
      f(R)+1&>&0~,\quad F(R)>0~, 
      \\ \label{condi2}
      r'&\ge& 0~,\quad F'(R)\ge 0~.
    \end{eqnarray}
    Eqs.(\ref{condi1}) are necessary to make $\exp(\lambda)$ and $r$
    positive, whereas Eqs.(\ref{condi2}) exclude crossings of
    different spherical shells of the collapsing matter, which
    simplifies discussion.
    
    Each shell of matter is characterized by some
    value of $R$, which remains constant during the collapse
    \begin{eqnarray}
      \label{R=c}
      R=\mathrm{const}~.
    \end{eqnarray}
    Integrating Eq.(\ref{rdot}) when Eq.(\ref{R=c}) holds one finds
    the explicit relation that defines the radius vector $r=r(R,\tau)$
    of the shell of matter
    \begin{eqnarray}
      \label{tau}
      \tau = -\int^r_{r_0} \frac{dr}{(f+F/r)^{1/2}}~.
    \end{eqnarray}
    Here $r_0=r_0(R)$ is an arbitrary function of $R$. The sign minus
    in front of the integral in Eq.(\ref{tau}) indicates that all
    velocities of the collapsing matter are negative, $\dot{r}<0$.  A
    fixed value of $R$ in Eq.(\ref{tau}) prompts a shortcut notation
    $f=f(R)$, $F=F(R)$, which is also used below. If one changes the
    sign in front of the integral in Eq.(\ref{tau}), then the metric
    in Eq.(\ref{metric}) describes the exploding matter i.e. the white
    hole, which plays a significant role below.
       
    In the particular case, when 
    \begin{eqnarray}
      \label{f=0}
      f(R)=0~,\quad F(R)=r_g=\mathrm{const}~, 
   \end{eqnarray}
   the Tolman metric Eq.(\ref{metric}) represents the metric of
   eternal black holes. To see this clearly, one can choose
   $r_0=r_g[3R/(2r_g)]^{3/2}$, finding that
   $r=r_g[3/(2r_g)(R-\tau)]^{3/2}$. Then Eq.(\ref{metric})
   reduces to the Lemaitre metric
    \begin{eqnarray}
      \label{lem}
      ds^2=-d\tau^2 +(r_g/r)\,dR^2 +r^2 d\Omega^2~.
    \end{eqnarray}
    Further transformation of variables $R,\tau \rightarrow r,t$
    \begin{eqnarray}
      \label{tautr}
      \!\!\!\!\tau = \pm t \pm \int \!\frac{(r_g/r)^{1/2}}{1-r_g/r}dr,~\,
      R = r + \int \!\frac{(r/r_g)^{1/2}}{1-r_g/r}dr,
    \end{eqnarray}
    converts the Lemaitre metric to the Schwarzschild metric
    \begin{eqnarray}
      \label{schw}
      ds^2=
    -\left(1-\frac{r_g}{r} \right)dt^2 
    +\frac{dr^2}{1-r_g/r }+r^2 d\Omega^2\,,
    \end{eqnarray}
    which explicitly describes a static eternal black hole.
    
    \section{Complex transformations and time inversion for collapsing
      matter}
    \label{riemann}
    Let us introduce a complex structure associated with the Tolman
    metric Eq.(\ref{metric}). The angular variables $\theta,\varphi$
    would not play a role below because we restrict discussion to the
    spherically symmetrical case. Our aim is to generalize the radial
    motion, allowing the radial variables to acquire complex values.
    It is sufficient to keep $R$ real, allowing only $r$ and $\tau$ to
    become complex-valued.  It suffices also to assume that the
    classical equations of motion for matter
    Eqs.(\ref{R=c}),(\ref{tau}) hold; in other words, we consider here
    the analytical continuation of the geodesics for the collapsing
    matter.
    
\begin{figure}[tbh]
      \centering
      \includegraphics[height=4cm,keepaspectratio=true]{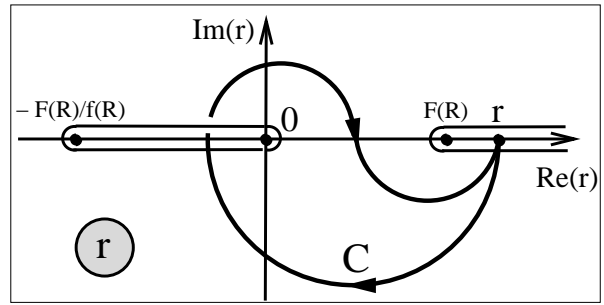}
      \caption{The Riemann surface for the
        function $\tau(r)$ possesses the cut of the squareroot nature
        that connects the points $r=0$ and $r=-F/f$. Similar cut
        exists for the action $S(r,\tau(r))$ of the probing particle.
        Additionally, the action has the logarithmic singularity on
        the horizon $r= F(R)$ with the corresponding cut attached. The
        analytical continuation along the contour $C$ generates the
        time inversion Eq.(\ref{t-t}) for the collapsing matter;
        additionally, it results in transformations
        Eqs.(\ref{Sin-out}),(\ref{Sout-in}) for the action of the
        probing particle.}
      \label{one}
    \end{figure}

    Eq.(\ref{tau}) allows one to take one of the variables $r,\tau$ as
    an independent variable, and treat the other one as its function.
    The conventional approach (which was followed above) takes $r$ as
    a function of $\tau$.  However, it is more convenient here to
    assume that $\tau$ is a function of $r$, $\tau=\tau(r)$, see Fig.
    \ref{one}.  This choice allows one to keep connection with the
    case of eternal static black holes, where the time variable does
    not play a role and the analytical continuation is applied to $r$,
    see Section \ref{et}.
    Obviously the function $\tau(r)$ presented by the integral in
    Eq.(\ref{tau}) can be expressed via elementary functions, but it
    is more instructive to derive its analytical properties directly
    from Eq.(\ref{tau}). On the complex plane $r$ the integral in
    Eq.(\ref{tau}) has two singular points, $r=0$ and $r=-F/f$, which
    match singularities of the radical $(f+F/r)^{1/2}$ in the
    integrand. This makes $\tau(r)$ an analytical function of $r$,
    which has a cut on the complex plane $r$ that runs between the two
    singular points $r=0$ and $r=-F/f$. Fig.  \ref{one} shows this cut
    running over the section $(-F/f,0)$ of the real axis.  Using the
    analytical continuation over this cut one constructs the Riemann
    surface for this function.  The singularities at $r=0$ and
    $r=-F/f$ are of the squareroot nature.  Therefore the Riemann
    surface includes two Riemann sheets.
    
    Let us assume that the initial real physical values of $r$ and
    $\tau$ belong to the outside region, i. e. $r>=F(R)$, where
    $R=R(r,\tau)$. Starting from this point one can continue the
    function $\tau(r)$ along the contour $C$ that crosses the cut and
    then returns to the initial physical value of $r$, as shown in
    Fig.  \ref{one}. Under this transformation (usually called a
    monodromy) the function is transformed as
    \begin{eqnarray}
      \label{t-t}
      \tau(r)\rightarrow -\tau(r)+\alpha~.
    \end{eqnarray}
    Here the sign minus arises due to the squareroot nature of the
    crossed cut.  The term $\alpha$ in Eq.(\ref{t-t}) is a constant,
    which depends on $r_0$ in Eq.(\ref{tau}). Since $r_0$ is real,
    $\alpha$ is also real and can be absorbed in the shift of the
    origin of time.
    
    Eq.(\ref{t-t}) shows that the analytical continuation described
    results in the inversion of the time variable. The equations of
    motion of the collapsing matter under this continuation become
    equations of motion for the expanding matter.  Speaking plainly,
    this analytical continuation transforms a black hole into a
    white hole.

    \section{Probing particle}
    \label{proba}
    Consider motion of a probing particle in the metric
    Eq.(\ref{metric}) created by the collapsing matter. It suffices to
    restrict discussion to the simplest case of a pure radial motion
    of the massless scalar particle, i.e. to choose the spin $s$,
    orbital momentum $l$, and mass $\mu$ of the particle to be all
    zero, $s=l=\mu=0$.  The corresponding classical action is a
    function of $R$ and $\tau$ only, $S=S(R,\tau)$.  The
    Hamilton-Jacobi equations of motion $g^{\mu\nu}\partial_\mu S \,
    \partial_\nu S=0$, which follow from Eq.  (\ref{metric}) in this
    case, have a simple form
    \begin{eqnarray}
      \label{tauR}
      \frac{\partial S}{\partial \tau} 
      =\pm\,\frac{1}{r'} \,(f+1)^{1/2}  \,
      \frac{\partial S}{\partial R}~.
    \end{eqnarray}
    Here $f=f(R)$; $r$ is defined in Eq.(\ref{tau}) as a function of
    $R$ and $\tau$, $r=r(R,\tau)$; the signs plus and minus correspond
    to the outgoing and incoming trajectories respectively.
    Eq.(\ref{tauR}) is formulated in terms of the initial variables
    $R,\tau$ of the Tolman metric.  It is convenient now to look at
    the problem from another perspective, considering $r,\tau$ as
    independent variables and assuming that $R=R(r,\tau)$ is a
    function defined by the geodesics Eq.  (\ref{tau}). One easily
    finds that
    \begin{eqnarray}
      \label{part1}
      \left( \frac{\partial S}{\partial R} \right)_\tau &=&
           r' \left( \frac{\partial S}{\partial r} \right)_\tau~,
           \\ \label{part2}
      \left( \frac{\partial S}{\partial \tau} \right)_R &=&
           \left( \frac{\partial S}{\partial \tau} \right)_r-r'\dot{R}
      \left( \frac{\partial S}{\partial r} \right)_\tau~,
    \end{eqnarray}
    where $\dot{R} =(\partial R/\partial \tau)_r$. Differentiating
    Eq.(\ref{tau}), while keeping either $\tau =\mathrm{const}$, or
    $r=\mathrm{const}$, one finds also that
    \begin{eqnarray}
      \label{rR}
      r'\dot{R} = (f+F/r)^{1/2}~.
    \end{eqnarray}
    Using Eqs.(\ref{part1}),(\ref{part2}) and (\ref{rR}) one rewrites
    Eq.(\ref{tauR}) in a more convenient form
    \begin{eqnarray}
      \label{tr}
      -\frac{\partial S}{\partial \tau}= w
      \frac{\partial S}{\partial r}~,
    \end{eqnarray}
    where $S=S(r,\tau)$ and the function $w$ is defined as
    \begin{eqnarray}
      \label{w}
      w \equiv w_\pm(r,\tau) = -\big( \, (f+F/r)^{1/2}\mp\,(f+1)^{1/2}\,\big)~.
    \end{eqnarray}
    The signs plus and minus here match the signs in Eq.(\ref{tauR}),
    the variables $f=f(R)$ and $F=F(R)$ are both functions of $r,\tau$
    since it is presumed that $R=R(r,\tau)$.
    
    To clarify the physical meaning of the factor $w$ in Eq.(\ref{tr})
    it is convenient to use the canonical formalism introducing the
    Hamiltonian $H$ and the momentum of the particle $p$
    \begin{eqnarray}
      \label{mom}
      p&=& \frac{\partial S}{\partial r}~,
      \\ 
      \label{ham}
      H&=&-\frac{\partial S}{\partial \tau}=wp~,
    \end{eqnarray}
    where the last identity follows from Eq.(\ref{tr}). Then from
    Eq.(\ref{ham}) and the canonical equations of motion one finds
    that $w$ equals the velocity of the particle
    \begin{eqnarray}
      \label{canon}
      \frac{dr}{d\tau}=\frac{ \partial H}{\partial p}=w~,
    \end{eqnarray}
    which makes Eq.(\ref{tr}) convenient for the following discussion.

    \section{Complex transformations for probing particle}
    \label{collapsar}
       
    Let us discuss an analytical continuation of the variables, which
    describe the radial motion with zero orbital momentum of the
    scalar massless probing particle. This task is similar to the one
    dealt with in Section \ref{riemann}, where the analytical
    continuation was applied to the collapsing matter.  However the
    study of the probing particle will be more detailed. In Section
    \ref{riemann} we discussed the analytical continuation assuming
    that the equations of motion are satisfied (i.e. assuming that
    Eqs.(\ref{R=c}),(\ref{tau}) are valid). For the probing particle
    discussed here the analytical continuation of the action of the
    probing particle is considered. This presumes that the classical
    equations of motion for this particle are not necessarily valid.
    The classical action presents a particular interest since it
    allows one to construct the semi-classical wave function, see
    Section \ref{wave}.
    
    Since the variables $r,\tau$ are not presumed to satisfy the
    equations of motion, the action $S(r,\tau)$ that describes the
    radial motion is a function of two independent variables. An
    analytical structure associated with this function for complex
    values of both variables is too sophisticated for our purposes.
    It is desirable therefore to simplify the problem forcing the
    action to be a function of only one complex variable.
    
    With this in mind let us adopt the following approach.  First,
    take some arbitrary real physical values for $r,\tau$, call them
    $r=r_0$ and $\tau =\tau_0$. Then find a real physical $R$ that
    satisfies classical equations of motion for the collapsing matter
    Eqs.(\ref{R=c}),(\ref{tau}). In other words, find a shell of
    matter labeled by $R$, which at the moment $\tau_0$ is located at
    the radius-vector $r_0$. This makes $R$ a function of the chosen
    physical variables $R=R(r_0,\tau_0)$.  After that let us fix the
    found value of $R$ and allow $r$ and $\tau$ to take only
    particular complex values that satisfy Eq.(\ref{tau}).
    
    Thus, the described procedure presumes that the complex values of
    $r,\tau$ satisfy the classical equations of motion (\ref{tau}) for
    matter. The complex radius vector $r$ of the particle follows the
    complex extension of the geodesics.  Under this condition the
    complex-valued $\tau$ can be taken as a function of the
    complex-valued $r$, $\tau=\tau(r)$.  At the same time, there is no
    restriction on real physical values of $r_0,\tau_0$; they remain
    two independent physical variables.  The procedure outlined
    ensures that during the analytical continuation the action remains
    a function of only one complex variable $r$,
    $S(r,\tau)=S(r,\tau(r))$, which simplifies its analytical
    properties.
        
    There exist, obviously, a variety of other ways to put some
    restriction on complex values of $r$ and $\tau$ making $\tau$ a
    function of $r$, and thus forcing the action to be a function of
    one variable.  An advantage of the procedure adopted here is that
    it matches the motion of matter, which follows the geodesics. This
    fact will allows one to combine results for the analytical
    continuation of the motion of the matter and probing particle.
    
    It becomes clear below that the horizon plays an important role in
    the analytical structure of the action. Anticipating this feature
    one can presume that the physical values of $r_0,\tau_0$ are chosen
    close to the event horizon, i. e. the following condition is
    fulfilled
    \begin{eqnarray}
      \label{hori}
      |r_0-F(R)| \ll F(R)~,
    \end{eqnarray}
    where $R$ is a function of $r_0,\tau_0$ as described above. This
    means that at the moment $\tau_0$ the shell marked by $R$ is close
    to its horizon as well. The subscripts in $r_0,\tau_0$ are
    suppressed below; it is to be clear from the context whether real
    physical values of the variables $r,\tau$, or their complex
    extensions are discussed.
        
    Eq.(\ref{w}) defines $w$ as a function of $r,\tau$.  Following the
    procedure adopted, i. e. taking $R=\mathrm{const}$, one makes $w$
    a function of $r$ only. Then Eq.(\ref{w}) shows explicitly that on
    the complex plain $r$ this is a regular function with a
    squareroot-type cut from $r=0$ to $r=-F/f$. This function takes
    the values $w_+$ and $w_-$ on the two sheets of the corresponding
    Riemann surface. It is convenient to label these sheets as $A_+$
    and $A_-$, and call them the sheet of outgoing particles, and the
    sheet of incoming particles respectively (keeping in mind that the
    velocities satisfy $w_+>0$, $w_-<0$).
    
    Since the action $S(r,\tau)$ is a solution of Eq.(\ref{tr}), which
    is governed by $w$, it necessarily exhibits properties similar to
    $w$. This means that it is an analytical function, which possesses
    the cut of the squareroot nature from $r=0$ to $r=-F/f$ on the
    complex plane $r$, taking the values $S_+(r,\tau(r))$ and
    $S_-(r,\tau(r))$ on the two sheets of the Riemann surface.  Here
    $S_+(r,\tau)$ is defined on the sheet $A_+$ and describes the
    outgoing motion, while $S_-(r,\tau)$ is located on $A_-$ and is
    related to the incoming motion of the particle.
    
    Let us discuss now the second important feature of the action,
    which is related to the node of the function $w$.  Eq.(\ref{w})
    shows that this node is located on the horizon $r=F$ on the sheet
    $A_+$ of the Riemann surface, which corresponds to the outgoing
    motion with $w=w_+$.  This singularity of Eq.(\ref{tr}) enforces
    singular behavior on the action. The derivative $-\partial S /
    \partial \tau=H$ on the left-hand side of Eq.  (\ref{tr})
    represents the Hamiltonian, which is certainly positive $H>0$ for
    all physical values of $r, \tau$ in the outside region $r>F$.  It
    remains therefore nonnegative on the horizon $H \rightarrow H_0
    \ge 0, ~r\rightarrow F$.  One concludes that the derivative
    $\partial S/\partial r$ on the right-hand side of Eq.  (\ref{tr})
    exhibits a pole-type behavior on the horizon
    \begin{eqnarray}
      \label{pole}
      \frac{\partial S}{\partial r } =\frac{H_0}{w} \rightarrow
      \frac{a}{r-F}+\mathrm{const}~,
      \quad \mathrm{when}\quad r\rightarrow F~,
    \end{eqnarray}
    Here the residue is
    \begin{eqnarray}
      \label{aaa}
      a = 2\,(f+1)^{1/2}\,F \,H_0\ge 0~,
    \end{eqnarray}
    as it follows from Eq.(\ref{w}). The factor $H_0=-(\partial
    S/\partial \tau)_{r=F}$ in Eq.(\ref{pole}) denotes the Hamiltonian
    of the particle on the horizon. From Eq.(\ref{pole}) one finds
    that the action itself has a logarithmic singularity on the
    horizon
    \begin{eqnarray}
      \label{log}
      S  \rightarrow a \, \ln(r-F)+\mathrm{const}~,
      \quad \quad r\rightarrow F~.
    \end{eqnarray}
    We see that the analytical structure of the function
    $S(r,\tau(r))$ on the complex plane $r$ is defined by two features
    shown in Fig.  \ref{one}. There is a cut from $r=0$ to $r=-F/f$
    that connects the two sheets $A_+$ and $A_-$ of the Riemann
    surface where the action takes the values $S_+(r,\tau)$ and
    $S_-(r,\tau)$ respectively.  Additionally, there is the
    logarithmic singularity Eq.(\ref{log}), which is located on the
    horizon being present on the sheet $A_+$ \footnote{The cut
      attached to this logarithmic singularity gives access to an
      infinite set of additional sheets of the Riemann surface, though
      they will not play a role in the following discussion.}.
    
  Consider now the transformation of the action when it is continued
  along the contour $C$ in Fig. \ref{one}. The reasons inspiring to
  choose this contour for the analytical continuation are discussed
  below. To be specific let us take the action that describes the
  incoming trajectory $S_-=S_-(r,\tau(r))$.  Correspondingly the
  contour $C$ starts on the sheet $A_-$ of the Riemann surface that
  corresponds to the incoming particles and, after crossing the cut
  over $(-F/f,0)$, ends up on the sheet $A_+$ of the outgoing
  particles.
    
    Combining the transformations arising from both the cut on the
    segment $(-F/f,0)$ and the logarithmic singular point $r=F$, see
    Fig.  \ref{one}, one concludes that the analytical continuation
    over the contour $C$ results in the following transformation of
    the action
    \begin{eqnarray}
      \label{Sin-out}
      S_- (r,\tau) \rightarrow - S_+(r,\tau) + i \pi a ~.
    \end{eqnarray}
    Two features are to be noted here. First, the negative sign in
    front of $S_+(r,\tau)$, which is related to the crossing of the
    cut; the squareroot nature of this cut makes this sign inevitable.
    Second, there is the imaginary part in the right-hand side, which
    arises due to circling around the logarithmic singular point over
    the angle $\pi$ counter clockwise.  It adds $i \pi$ to the
    logarithmic function, forcing through Eq.(\ref{log}) the imaginary
    constant $i \pi a$ to appear in the transformation of the action.
    Generally speaking, there may exist also an additional real
    additive constant in the right-hand side of Eq.(\ref{Sin-out}),
    but its calculation is not pursued here, since it is not important
    for our purposes. In contrast, the imaginary term plays a
    prominent role below.
    
    Similarly the analytical continuation over the contour $C$ for the
    action $S_+(r,\tau)$ results in the transformation
    \begin{eqnarray}
      \label{Sout-in}
      S_+(r,\tau) \rightarrow -S_-(r,\tau)  - i\pi a~.
    \end{eqnarray}
    The negative sign in front of the imaginary part is due to the fact that
    the logarithmic singularity in this example is circled in the
    clockwise direction.
    
    In Section \ref{riemann} the analytical continuation that
    transforms the equations of motion for collapsing matter into
    equations of motion of the expanding matter was found. The results
    of the present Section are in line with this finding. The actions
    for incoming and outgoing particles transform one into another in
    Eqs.(\ref{Sin-out}),(\ref{Sout-in}).

    \section{particle in the field of eternal black hole}
    \label{et}
    It is instructive to specify the approach of Section
    \ref{collapsar} for the simplest possible case, when both
    functions $f,F$, which govern the collapse are constants
    \begin{eqnarray}
      \label{con}
      f=\mathrm{const}~,\quad F=r_g=\mathrm{const}~. 
    \end{eqnarray}
    This condition, which incorporates the physically interesting case
    of the eternal Schwarzschild black holes
    Eqs.(\ref{f=0}),(\ref{schw}), indicates that $w$ defined by
    Eq.(\ref{w}) is a function of only one variable $r$, which allows
    to integrate Eq.(\ref{tr}) explicitly
    \begin{eqnarray}
      \label{int}
      S_\pm(r,\tau) =
      -\int^r\frac{\varepsilon\,dr}{(f+F/r)^{1/2}\mp
      (f+1)^{1/2}}-\varepsilon \tau~.
    \end{eqnarray}
    Here $\varepsilon$ is the energy of the particle. The sign minus
    in front of the integral in Eq.(\ref{int}) is related to negative
    velocities of the collapsing matter $\dot{r}<0$, compare
    Eq.(\ref{tau}).
       
    The explicit integral representation Eq.(\ref{int}) allows one to
    verify that the action possesses the analytical structure shown in
    Fig. \ref{one}. Namely, on the complex plane $r$ there exists a
    cut from $r=-F/f$ to $r=0$, which connects the two sheets $A_+$
    and $A_-$ of the Riemann surface. On the sheet $A_+$ there exists
    an additional logarithmic singularity. It arises from the node in
    the integrand in Eq.(\ref{int}), being located on the horizon
    \begin{eqnarray}
      \label{hora}
      r=F \equiv r_g~.
    \end{eqnarray}
    These facts confirm what is known from a more general approach of
    Section \ref{collapsar}. One also re-derives
    Eqs.(\ref{Sin-out}),(\ref{Sout-in}) specifying that for eternal
    black holes $a$ in Eq.(\ref{aaa}) is a constant
     \begin{eqnarray}
      \label{gH}
      a=2\pi r_g \varepsilon=\frac{\varepsilon}{2T_\mathrm{H}}~,
    \end{eqnarray}
    where 
    \begin{eqnarray}
      \label{TH}
      T_\mathrm{H}=\frac{1}{4\pi r_g}
    \end{eqnarray}
    is the Hawking temperature.

    \section{wave function}
    \label{wave}
    Consider the scalar massless particle that moves with the zero
    orbital moment in the gravitational field of a collapsar,
    describing its propagation in the semiclassical approximation with
    the help of the wave function
    \begin{eqnarray}
      \label{psi}
      \psi(r,t)=C\Big( \exp \big(i S_-(r,t)\,\big) + {\cal R}
      \exp \big(i S_+(r,t)\big)\Big).
    \end{eqnarray}
    Here $S_\mp(r,t)$ are the two actions that are related to the
    incoming and outgoing trajectories. Correspondingly, the first
    term in Eq.(\ref{psi}) describes the incoming wave, while the
    second one represents the outgoing wave.
    
    The wave function Eq.(\ref{psi}) can be used to describe the
    process of absorption of particles by the black hole. In this case
    one needs to consider the behavior of this wave function on the
    horizon of each shell, i.e. to presume that $r,\tau$ satisfy
    condition
    \begin{eqnarray}
      \label{rFF}
      |\,r-F(R)\,|\ll F(R)~, 
    \end{eqnarray}
    (which is similar to Eq.(\ref{hori})). Eq.(\ref{rFF}) has a clear
    physical meaning. It states that there is a shell of matter
    characterized by some particular value of $R=R(r,\tau)$, which at
    the moment of time $\tau$ has the radius vector $r$ that is close
    to the horizon of this shell $F(R)$. Same can be rephrased as a
    statement that the amount of dust matter that is accumulated
    inside the sphere of radius $r$ at the moment $\tau$ is close to a
    critical value, which is necessary to create a black hole.
    Generically, as the time passes on each shell converges to its own
    horizon $F(R)$ (eventually crossing it).  Therefore there exists a
    set of critical values $r_\mathrm{c}$ of the radius vector that
    are located in an interval $0<r_\mathrm{c}\le F_\mathrm{max},
    ~F_\mathrm{max}=\lim_{R\rightarrow \infty}F(R)$; if $r$ is close
    to any such critical value, then one can always choose $\tau$ to
    to satisfy Eq.(\ref{rFF}).  For static black holes the set of
    critical radiuses degenerates to the only value of the
    Schwarzschild radius $r_c=r_g$.
    
    Conventional wisdom prompts one to assume that in the vicinity of
    the horizon there should be no reflected wave; whatever comes
    close to the horizon must be absorbed, simply because we deal with
    a black hole, the object that is presumed to be {\it absolutely}
    black.  Thus, one could anticipate that the second term in
    Eq.(\ref{psi}) is absent, i.e. the factor $\mathcal{R}$, which we
    will call the reflection coefficient, is zero, $\mathcal{R}=0$
    \footnote{Eq.(\ref{psi}) is written in the semiclassical
      approximation. In regions where the semiclassical approach fails
      the second term in the wave function is definitely present,
      describing the flux of particles scattered by those parts of the
      gravitational potential that violate the semiclassical
      conditions. In relation to the Hawking radiation this phenomenon
      is often referred to as the grey body effect; for low energy
      particles it takes place at large distances from a black hole.
      In contrast, in the close vicinity of the horizon one should
      expect the semiclassical method to work well (as can be easily
      verified for eternal black holes). The appearance of the second
      term here would indicate the reflection on the horizon.}.  The
    absence of the reflected wave on the horizon, often called the
    Matzner condition in the stationary problems, see the references
    in the book \cite{fullerman_handler_matzner_88}, played a
    fundamental role in the scattering problem; it was used in the
    pioneering Refs.
    \cite{press_teukolsky_72,press_teukolsky_74,starobinsky_73,
      starobinsky_churilov_73,unruh_76} as well as in numerous
    subsequent developments related to the scattering problem.
    
    It turns out, however, that the reflection coefficient proves be
    nonzero $|\mathcal{R}|>0$, the second term is present in
    Eq.(\ref{psi}) on the horizon. In other words there is the RH, as
    was shown in \cite{kuchiev_1,kuchiev_2,kuchiev_3} for eternal
    black holes.
    
    In order to verify this claim for the collapsing case let us apply
    the method outlined in Section \ref{main}, which culminates in
    Eq.(\ref{psipsi}) for the wave function $\Psi(Q)$ of the quantum
    system. The system considered here includes the collapsing dust
    matter plus the incoming probing particle. However, quantum
    treatment is necessary only for the particle, as the dust matter
    is much less affected by quantum effects.  This happens because
    the quantum effects prove strong only for low energies of a
    particle that approaches the horizon (see Eq.(\ref{R}) below),
    while each drop of the collapsing dust can be considered as a
    heavy particle that has high energy associated with its mass.
    Anticipating this fact from the very beginning, one can describe
    the matter in the pure classical approximation, taking quantum
    effects into account only in the wave function of the particle
    $\psi(r,t)$.
    
    Thus, Eq.(\ref{psipsi}) needs to be reformulated in terms of the
    wave function $\psi(r,t)$ of the particle, which moves in the
    given time-dependent metric produced by the collapsing matter.  As
    the first step in this direction, let us use the contour $C$ shown
    in Fig. \ref{one} for the analytical continuation. Firstly, using
    this contour one continues analytically the equations of motion
    Eqs.(\ref{R=c}),(\ref{tau}) for each shell $R$ of the dust matter,
    converting the equations for the collapsing matter into equations
    of the expanding matter (converting the black hole into the white
    hole). The way to do this was described in detail in Section
    \ref{riemann}.  Secondly, using same contour $C$ one can fulfill
    the necessary transformation of the wave function of the particle.
    Using Eqs.(\ref{Sin-out}),(\ref{Sout-in}) for the classical action
    one finds that this operation can be written as a transformation
    of the wave function $\psi(r,t) \rightarrow \tilde \psi(r,t)$,
    where
    \begin{eqnarray}
      \label{psi1}
      \tilde \psi(r,t) = Ce^{i\gamma} \Big( \rho \exp \big( -i
      S_-(r,t)\,\big) 
      \\ \nonumber + 
      \frac{\cal R}{\rho}e^{i\delta}
      \exp \big(-i S_+(r,t)\,\big)\Big)~.
    \end{eqnarray}
    Here $\rho$ is governed by the imaginary constant in the
    transformation of the action (\ref{Sin-out}),(\ref{Sout-in})
    \begin{eqnarray}
      \label{rho}
      \rho = \exp(-\pi a)~,
    \end{eqnarray}
    where $a$ is defined in Eq.(\ref{aaa}), while the phases
    $\gamma,\delta$ are related to the real constants in the
    transformation of the action, which were omitted in
    (\ref{Sin-out}),(\ref{Sout-in}). Clearly, the first main term in
    Eq.(\ref{psi1}) describes the outgoing wave, indicating that this
    wave function describes the creation of the particle. This is a
    sensible result since the analytical continuation transforms the
    matter into the white hole, as discussed above. The white whole is
    able to create particles.  Thus, the considered analytical
    continuation leads to the time inversion.
    
    Since the time invariance is valid, one is certainly free to
    fulfill the inversion of time directly in the equations of motion.
    By applying the inversion of time to the classical equations of
    motion for the dust matter one brings them back to the collapsing
    case.  The inversion of time leads also to the complex conjugation
    of the wave function of the particle, i.e. to the transformation
    \begin{eqnarray}
      \label{psi2}
      \psi(r,t) \rightarrow \psi'(r,t)  = \psi^*(r,t)~.
    \end{eqnarray}
    Thus the inversion of time can be fulfilled in two different ways.
    either by the analytical continuation, resulting in Eq.
    (\ref{psi1}), or it can be done directly in the equations of
    motion, which leads to Eq.(\ref{psi2}). Since these two ways
    should give the same physical result, the wave functions in
    Eqs.(\ref{psi1}),(\ref{psi2}) should be the same, or at worst
    differ by some phase factor.  One derives from this 
    \begin{eqnarray}
      \label{sym}
      \tilde \psi(r,t) = e^{i\chi}\,\psi'(r,t)=e^{i\chi}\,\psi^*(r,t)~.
    \end{eqnarray}
    This equality presents a restriction on the wave function of the
    particle, which stems from the symmetry related to the time
    inversion (compare Eq.(\ref{psipsi}), which was written in general
    terms).  From definitions of the wave functions Eqs.
    (\ref{psi}),(\ref{psi1}) and Eq.(\ref{sym}) one derives that the
    reflection coefficient is necessarily nonzero
    \begin{eqnarray}
      \label{R}
      |\mathcal{R}|= \rho = \exp(-\pi a) > 0~.
    \end{eqnarray}
    Thus, the invariance under the time inversion Eq.(\ref{sym}) makes
    the process of the RH inevitable, the horizon is able to reflect
    the incoming particles.
    
    Derivation of Eq.(\ref{sym}) was based on a particular contour on
    the complex plane $r$ shown in Fig. \ref{one}. This contour must
    necessarily cross the cut that connects the sheets $A_+$ and $A_-$
    of the Riemann surface, thus leading to the time inversion in the
    equations of motion of the matter Eq.(\ref{t-t}) and
    transformation of the wave function Eq.(\ref{psi1}).  The chosen
    way of curling around the logarithmic singularity leads to a
    sensible result, $|\mathcal{R}|<1$ in Eq.(\ref{R}), while the
    opposite direction of circling around this singularity would
    result in the violation of unitarity $|\mathcal{R}|>1$. Such
    dependence of the result on the contour is common in semiclassical
    problems, see the book \cite{landau_lifshits_III}. 
     
    Eq.(\ref{R}) shows that the RH really takes place, the horizon
    reflects particles.

     \section{Discussion}
     \label{dis}
     
     \subsection{Main result}
     
     Our main result is presented in Eqs.(\ref{psi}),(\ref{R}), which
     show that the wave function of the particle approaching the black
     hole horizon has a component that describes the reflected wave.
     In other words, the particle can bounce on the horizon back into
     the outside world.  According to Eq.(\ref{R}) the probability of
     this effect equals
     \begin{eqnarray}
       \label{P}
       P =|\mathcal{R}|^2= \exp(-2\pi a) > 0~,
     \end{eqnarray}
     where the parameter $a$ is expressed in Eq.(\ref{aaa}) via the
     functions $f=f(R),~F=F(R)$, which define the collapse. Their
     argument $R=R(r,\tau)$ depends on the coordinates $r,\tau$ of the
     wave function. Additionally, $a$ depends on the Hamiltonian of
     the particle $H_0=-\partial S/\partial \tau$, which is taken on
     the horizon.  The expression for $a$ simplifies in the limit of
     the static Schwarzschild black hole, $f(R)\rightarrow
     0,~F(R)\rightarrow r_g=\mathrm{const}$ when Eq.(\ref{gH}) is
     valid.  In this case Eq.(\ref{P}) reduces to
     \begin{eqnarray}
       \label{PS}
       P = \exp(-\varepsilon/T_\mathrm{H}) > 0~,
     \end{eqnarray}
     where $T_\mathrm{H}$ is the Hawking temperature Eq.(\ref{TH}).
     Eq.(\ref{PS}) coincides with the result of
     \cite{kuchiev_1,kuchiev_2,kuchiev_3} derived exclusively for the
     static case. Eq.(\ref{PS}) shows that the reflection is strong
     for low energy particles $\varepsilon < T_\mathrm{H}$. In other
     words, in the low energy limit the horizons of black holes act as
     mirrors.

     \subsection{Symmetry condition}
     It is interesting to compare the symmetry condition
     Eq.(\ref{sym}) with a similar condition found for the stationary
     case in \cite{kuchiev_1,kuchiev_2,kuchiev_3}. The latter works
     focused on the wave function of the particle, leaving aside the
     effect of the discussed transformation on the collapsing matter.
     This simplified the consideration, but raised several qualitative
     questions.  First, the symmetry, as it was derived in
     \cite{kuchiev_1,kuchiev_2,kuchiev_3}, was formulated in terms of
     regions I and III of the Kruskal plane \cite{kruskal_1960}
     (definitions of regions follow
     \cite{misner_thorne_wheeler_1973}).  However, one could argue
     that the collapsing black hole moves from region I into II, and
     does not come across region III. It was unclear therefore whether
     the proposed symmetry condition would hold for collapsing black
     holes. Second, there are known several ways to define the
     topological correspondence between regions I and III, see Refs.
     \cite{gibbons_86,chamblin_gibbons_95}.  Refs.
     \cite{kuchiev_1,kuchiev_2,kuchiev_3} suggested to use a
     particular correspondence (the most obvious), postponing a
     question of whether the symmetry condition remains valid for all
     other possible ways of the topological correspondence.
     
     The symmetry transformation in the present work is discussed from
     a broader perspective (in addition to the fact that a more
     general collapsing case is considered). The inversion of the time
     is discussed in relation to both the wave function of the probing
     particle, and the motion of the matter. This point of view
     enables one to formulate simple physical answers for the
     mentioned problems.  First, the inversion of the time variable
     not only brings the particle to region III, but it also
     transforms the collapsing matter into the expanding white hole,
     thus making region III relevant to the matter as well. Second,
     for any sensible topological correspondence between regions I and
     III the operation of the time inversion remains a symmetry of the
     space-time. This symmetry guarantees that the presented approach
     remains applicable for any topological identification between
     these regions.
     
     \subsection{Relation to Aharonov-Bohm effect}
     
     The RH can be considered as a manifestation of the logarithmic
     singularity, which the action exhibits on the event horizon.  At
     the same time the classical trajectories do not come across this
     singularity (as one easily verifies using Fig. \ref{one}).
     Therefore the singularity has no influence on the classical
     equations of motion, being important only in the action.  This
     property of the problem has a similarity with the Aharonov-Bohm
     effect, as was mentioned in \cite{kuchiev_2,kuchiev_3} in
     relation to the case of eternal black holes. Earlier a similarity
     that exists between physical events in the vicinity of the
     horizon and the Aharonov-Bohm effect was discussed in
     Ref.\cite{parikh_wilczek_00}.

     \section{Conclusion}
     
     It is shown that the reflection on the horizon of black holes is
     a general phenomenon, which manifests itself for the collapsing
     case as well as for eternal black holes.  One of interesting
     consequences of this fact is related to the information paradox.
     Since any incoming wave can bounce on the horizon back into the
     outside world, the horizon cannot destroy completely all the
     incoming information, at least part of it inevitably returns into
     the outside world.
      
     This work was supported by the Australian Research Council.
     Discussions with V.V.Flambaum are appreciated.

\end{document}